# Effects of West Coast forest fire emissions on atmospheric environment: A coupled satellite and ground-based assessment

**Srikanta Sannigrahi[a*], Qi Zhang[b], Francesco Pilla[a], Bidroha Basu[a], Arunima Sarkar Basu[a]**

[a] School of Architecture, Planning and Environmental Policy, University College Dublin Richview, Clonskeagh, Dublin, D14 E099, Ireland.

[b] Frederick S. Pardee Center for the Study of the Longer-Range Future, Frederick S. Pardee School of Global Studies, Boston University, Boston, MA 02215, USA

*Corresponding author: **Srikanta Sannigrahi**
E-mail: (Srikanta Sannigrahi*) : srikanta.sannigrahi@ucd.ie

**Abstract**

Forest fires have a profound impact on the atmospheric environment and air quality across the ecosystems. The recent west coast forest fire in the United States of America (USA) has broken all the past records and caused severe environmental and public health burdens. As of middle September, nearly 6 million acres forest area were burned, and more than 25 casualties were reported so far[1]. In this study, both satellite and in-situ air pollution data were utilized to examine the effects of this unprecedented wildfire on the atmospheric environment. The spatial-temporal concentrations of total six air pollutants, i.e. carbon monoxide (CO), nitrogen dioxide ($NO_2$), sulfur dioxide ($SO_2$), ozone ($O_3$), particulate matter ($PM_{2.5}$ and $PM_{10}$), and aerosol index (AI), were measured for the periods of 15 August to 15 September for 2020 (fire year) and 2019 (reference year). The in-situ data-led measurements show that the highest increases in CO (ppm), $PM_{2.5}$, and $PM_{10}$ concentrations ($\mu g/m^3$) were clustered around the west coastal fire-prone states, during the 15 August - 15 September period. The average CO concentration (ppm) was increased most significantly in Oregon (1147.10), followed by Washington (812.76), and California (13.17). Meanwhile, the concentration ($\mu g/m^3$) in particulate matter (both $PM_{2.5}$ and $PM_{10}$), was increased in all three states affected severely by wildfires. Changes (positive) in both $PM_{2.5}$ and $PM_{10}$ were measured highest in Washington (45.83 and 88.47 for $PM_{2.5}$ and $PM_{10}$), followed by Oregon (41.99 and 62.75 for $PM_{2.5}$ and $PM_{10}$), and California (31.27 and 35.04 for $PM_{2.5}$ and $PM_{10}$). The average level of exposure to CO, $PM_{2.5}$, and $PM_{10}$ was also measured for all the three fire-prone states. The results of the exposure assessment revealed a strong tradeoff association between wildland fire and local/regional air quality standard. The devastating west coast fire demonstrates the adverse effect of wildfire on the air quality status of a region.

**Introduction**

The West Coastal states of the United States of America (USA) have witnessed the historically worst wildfires which have swiped away a substantial forest area (as of September 17, 2020, nearly 2.5 million acres were burned in California and more than 5 million acres forest areas burned in California, Oregon, and Washington[2]) (National Interagency Fire Center, 2020). Many observations came out to estimate the total loss attributed to these fire events. In the Washington state, the report stated that more than 330,000 acres (133,546 hectares) of forest area were burned in a 24-hour period, which is also recorded highest comparing the preceding year's fire incidents of the state[3] (Washington State Department of Natural Resources, 2020[4]). In California, wildfires in 2020 are 26 times higher than the forest area (acres) burned in 2019 during the same time period[5]. According to California Department of Forestry and Fire Protection (CAL FIRE[6], 2020), 5 of the top 20 largest wildfires in California's history have occurred in 2020. Meteorological factors, especially the wind speed recorded up to 75 mph during the fire seasons, amplified the spread of the fire to the nearby states that rarely experience such devastating and intense fire blazes in the recent years[7]. Apart from the wind, NASA report[8] stated several other confounding factors, including record-breaking temperature and unusually dry air, exceeding the flammability threshold, prolonged dry period, abundant load of dry fuel, lack of soil moisture, etc., which have also been played key roles in amplifying the fire intensity and exacerbated the burn severity across the regions (NASA Earth Observatory, 2020). Among the active fires detected in California, Washington, and Oregon states in 2020, the August Complex fire, in Tehama County, California, was found to be the biggest fire ever recorded, covering more than 750,000 acres (CAL FIRE, 2020). According to European Commission's Copernicus Atmosphere Monitoring Service (CAMS) Data[9], the wildfires in 2020 are recorded to be "significantly more intense" compared to the average fire incidents during 2003-2019 period and found worst in 18 years (CAMS, 2020). The economic loss attributed to these calamities could be more than 20 Billion $[10].

---

[1] https://www.nytimes.com/2020/09/15/us/oregon-fires-california.html
[2] https://www.nytimes.com/2020/09/15/us/oregon-fires-california.html
[3] https://news.sky.com/story/california-wildfires-several-people-dead-as-blazes-ravage-state-12067759
[4] https://www.dnr.wa.gov/Wildfires

Although wildfire has a few positive impacts, such as enhancing resilience and adaptability of an ecosystem to combat climate extremes, increasing soil fertility, and removing the dead organic residuals from soil substrate, (Verma and Jayakumar, 2012; Li et al., 2009), several researchers have found a data-driven detrimental impact of forest fires led emission of fine particulate, black carbon (smoke), and other gaseous compounds on human health (B. et al., 2011; Aditama, 2000; Cascio, 2018; Schweizer and Cisneros, 2017). The health risk of wildfire emission, especially emissions of carbon and particulate matter, are found highly prominent in Southern and Western parts of the USA (Fowler, 2003). According to Fowler (2003), there are several factors, such as fuel type, fire behaviours, wind conditions, fire intensity, fire temperature, and fuel conditions (types of biomass and moisture levels in soils), determining how human health would respond to the unprecedented fire extreme conditions. Fire intensity, the other critical component, often regulates the combustion and emission of wildfires (Fowler, 2003; Sommers et al., 2014). High-intensity fires mainly emit carbon and water content, while, the lower intensity wildfires characterized by incomplete combustion and therefore, emits harmful pollutants in high volume (Loehman et al., 2011; Fowler, 2003; Loehman et al., 2011). However, Johnson (1992) study found a contrasting nature of the association between wildfires and loss of soil carbon storage. This study (Johnson et al.) noted that low-intensity fire has a trivial impact on degrading soil carbon and resulted in emission, while, an intense wildfire can be attributed to substantial loss of soil carbon.

Many key air pollutants, i.e. fine and coarse particulate matter ($PM_{2.5}$, $PM_{10}$), carbon monoxide (CO), nitrogen oxides (NOx), nitrous oxide ($N_2O$), methane ($CH_4$), volatile organic carbon (VOC), and many other toxics gases that emit during wildland fires have a profound impact at worsening air quality across the region (Urbanski et al., 2008; Cascio, 2018). Among the major air pollutants, fine particulate matter is the most common pollutant that increases significantly during the wildfire (Phuleria et al., 2005). This addition of $PM_{2.5}$ and $PM_{10}$ into the air causes severe adverse health burdens (Sannigrahi et al., 2020a, Kumar et al., 2020). Many cause-specific morbidities, such as respiratory and cardiovascular health issues, are often attributed to the hazardous concentration of fine particulate matter (B. et al., 2011; Aditama, 2000; Cascio, 2018; Sapkota et al., 2005; Landguth et al., 2020; Matz et al., 2020). Approximately, landscape fire smoke caused 339,000 premature deaths annually (Johnston et al., 2012). Landguth et al. (2020) study have found a strong linkage between high daily average of $PM_{2.5}$ concentrations that mainly comes from wildfire emission and upsurges of influenza cases (projected 16% or 22% growth in influenza rate per 1 μg/m$^3$ increase in average daily summer $PM_{2.5}$), during the wildfire season in the mountain west region of the USA. Butt et al., (2020) noted that the successful prevention of wildfires would avert nearly 16 800 (95CI: 16 300–17 400) premature deaths and 641 000 (95CI: 551 900–741 300) disability-adjusted life years (DALYs) in South America, where nearly 26% of the avoided premature mortality is located within the Amazon Basin.

---

[5] https://www.bbc.com/news/world-us-canada-54180049
[6] https://www.fire.ca.gov/
[7] https://news.sky.com/story/california-wildfires-several-people-dead-as-blazes-ravage-state-12067759
[8] https://earthobservatory.nasa.gov/images/147277/historic-fires-devastate-the-us-pacific-coast
[9] https://atmosphere.copernicus.eu/
[10] https://www.bbc.com/news/world-us-canada-54180049

Forest fires, caused by both natural and anthropogenic interventions, have a serious impact on the atmospheric environment and air quality across ecosystems (Ferrare et al., 1990); (Bo et al., 2020); (Cheng et al., 1998); (Sapkota et al., 2005); (Sannigrahi et al., 2020b); Sannigrahi et al., 2020b). Zhu et al., (2018) examined the impact of smoke aerosols from Russian forest fires on the air pollution over Asia using the Flexible Particle dispersion (FLEXPART) Weather Research and Forecasting (WRF) model and found that due to the 2016 Siberia fire, approximately 60–300 μg/m$^3$, 40–250 μg/m$^3$ and 5–140 μg/m$^3$ smoke aerosol concentrations were entered in Central Asia, Mongolia and Northern China, respectively. Johnston et al., (2014) measured the relationship between forest fire smoke events and hospital emergency department (ED) attendances in Sydney during 1996–2007 and found that Smoke events were associated with an instantaneous upsurge in presentations for respiratory conditions and had lagged consequences on ischemic heart disease and heart failure. Lazaridis et al., (2008) examined the effects of forest fire emissions on atmospheric pollution in Greece and found that the wildfire emissions have contributed 50% to the $PM_{10}$ concentration over the region affected by wildfires. DeBell et al., (2004) evaluated the regional air pollution status of the northeastern USA and its association with 2002 forest fire in Quebec and Canada; they found that biomass burning led plumes caused to enhance the CO mixing ratios and particulate matter concentration throughout the east coast of the USA. All these studies collectively indicate a strong synergic association between wildfires and degradation of air quality across the ecosystems. The forest fire in California, Oregon and Washington in 2020 has broken all past records, and a substantial forest area has already been lost due to this unprecedented catastrophe. A thorough and comparative assessment is therefore needed to discuss the severe effects of this wildfire on air quality amidst the COVID-19 pandemic time.

## 2. Materials and methods

### *2.1 West Coast wildland fires in 2020*

August Complex (including Doe Fire) wildfire is the largest single wildfire event recorded in California in 2020. The complex fire includes 38 separate fires started on 16 August 2020 due to lighting and fell under Mendocino National Forest, CAL FIRE Mendocino Unit, and CAL FIRE Humboldt-Del Norte Unit. Till now, the August complex fire events has spread over seven counties, i.e. Mendocino, Humboldt, Trinity, Tehama, Glenn, Lake, Colusa and destroyed 1017546 acres (South Zone 546,365 acres, North East Zone 267,115 acres, North West Zone 68,200 acres, West Zone 135,866 acres) of forest area (CAL FIRE, 2020). The complex wildfires include multiple fires. i.e. Elkhorn, Hopkins, Willow, Vinegar, and Doe fires. On September 9 2020, the Doe fire event, which was the main fire event of August complex, has surpassed the 2018 Mendocino Complex and become the largest single fire events in California. As of 8 October 2020, 210 structures damaged (residential, commercial and other) and 1 injury (confirmed fire personnel and civilian injuries) is recorded. Considering the immense size and devastating power of the complex fire event, the fire is being managed by three different zones, August Complex South Zone (Doe Fire), August Complex North Zone (Elkhorn Fire), and August Complex West Zone (western portion of Doe Fire and Elkhorn Fire). The other big fire events in California include SCU Lightning Complex (396,624 acres burnt and 100% contained as of 01/10/2020), LNU Lightning Complex (363220 acres and 100% contained as of 2/10/2020), North Complex Fire (still active, 318,930 acres burnt and 88% contained as of 9/10/2020), Creek Fire (still active, 330,899 acres burnt and 49% contained as of 8/10/2020), SQF Complex (still active, 164,993 acres burnt and 65% contained as of 8/10/2020), CZU Lightning (86509 acres and 100% contained as of 22/9/2020), El Dorado Fire (still active, 22,744 acres burnt and 95% contained as of 7/9/2020), Snow Fire (6013 acres and 15% contained as of 9/19/2020), and Shackleford Fire (50 acres and 90% contained as of 9/19/2020), are few of them (CAL FIRE, 2020). There are few other medium-small range fire events named as Foxfire, Bullfrog fire, Fork fire, Slater fire, Bobcat fire, Valley fire, Rattlesnake fire, Slink fire, Moraine fire, Dolan fire, Feather fire, Lake fire, and Apple fire, that occurred across California, Oregon, and Washington state (CAL FIRE, 2020). As of now, in California, a total of 4,040,935 acres of forest were burned, 8,320 wildfire incidents were recorded, 8 confirmed fatalities were identified, and 5,495 structures damaged or destroyed. The estimated acres burned due to forest fire in the previous years are recorded as 259,823 acres in 2019, 1,975,086 acres in 2018, 1,548,429 acres in 2017, 669,534 acres in 2016, 880,899 acres in 2015, 625,540 acres in 2014, 601,625 acres in 2013, respectively (CAL FIRE, 2020). According to CALFIRE daily wildfire incident report, five (August Complex Fire, Creek Fire, North Complex Fire, SCU Lightning Complex Fire, and LNU Lightning Complex Fire) of the top 20 biggest wildfires in California history have occurred in 2020. Additionally, since 2003, August Complex Fire is recorded to be the largest single wildfire events in California, followed by Mendocino Complex (July 2018, 459,123 acres burned and 280 structures damaged), SCU Lightning Complex (396,624 acres burned and 222 structures damaged), LNU Lightning Complex (363,220 acres burned and 1491 structures damaged).

## 2.2 Measuring air pollution status during the fire incidents period

### 2.2.1 Sentinel 5P TROPOMI upper atmosphere air pollution column density

Both satellite and in-situ air pollution data were utilized for quantitative assessment of the air quality status during the study period and subsequent interpretation. Sentinel 5P TROPOspheric Monitoring Instrument (TROPOMI) data was employed for measuring column number density of nitrogen dioxide ($NO_2$), carbon monoxide (CO), sulfur dioxide ($SO_2$), Ozone ($O_3$), and aerosol index (AI) for the 15 August to 15 September period for both 2019 and 2020 (Veefkind et al., 2012; Guanter et al., 2015; Kaplan et al., 2019; Sannigrahi et al., 2020b). The Sentinel 5P satellite mission is a single-payload satellite placed at low Earth orbit intended to monitor the concentration of trace gases and aerosols to provide crucial information about upper surface air quality, climate change, and depletion of Ozone layer. The playload of the Sentinel 5P mission is known as TROPOMI. TROPOMI is a spectrometer consists of multiple visible, ultraviolet, near-infrared, and shortwave near-infrared spectral bands at a spatial resolution as 7 km x 3.5 km (Veefkind et al., 2012). Sentinel 5P data was processed in Google Earth Engine cloud platform (Gorelick et al., 2017). The year 2020 was considered a fire year, and 2019 was considered as the reference year. The default unit of measurement ($mol/m^2$) of the satellite air pollution estimates was converted to concentration unit ($\mu g/m^3$) to make it comparable with the ground-based measurements. The Goddard Earth Observing System forward processing (GEOS-FP) model-based black carbon column mass density information was collected from NASA Erath Observatory. The Suomi NPP – Visible Infrared Imaging Radiometer Suite (VIIRS) 375m and National Oceanic and Atmospheric Administration (NOAA-20) VIIRS 375m active forest fire data was collected from Fire Information for Resource Management System (FIRMS[11], 2020). Sentinel 5P has been proven effective in detecting biomass burning emission (BBE) with the help of satellite-derived fire radiative power (FRP) (Li et al., 2020). FRP denotes to the radiative energy emitted from wildfire, has been effectively used as a reliable method for detecting BBE.

*2.2.2 In-situ air pollution*

In-situ air pollution measurements were retrieved from OpenAQ[12]. OpenAQ is a non-profit organization that aims to retrieve, harmonize, share open-air quality information across the citizens and organizations and to provide up-to-date status of clean air which would eventually help prevent air pollution led health burdens across the world. The OpenAQ platform retrieved latest and up-to-date air quality data from multiple sources including government/institution air quality monitoring stations, low-cost open-air quality sensors, etc. The current functionality that OpenAQ offers includes retrieving data using location and country information, through application programming interface (API). Till now, OpenAQ has 693M air quality measurement information, 150 data sources, 13000 locations, and covers nearly 95 countries. OpenAQ collects both public and organization hosted data based on five criteria: (1) data collection should consist any of the seven pollutants: $PM_{10}$, $PM_{2.5}$, $SO_2$, CO, $NO_2$, $O_3$, or black carbon (BC). These pollutants were considered as they are the most common air pollutants measured at outdoor environment. Though BC does not have any proven health effects, still, the same has been considered as it does have noted impact on regional and global climate warming; (2) data source must be hosted by or affiliated with official-level stationary (government entity or international organization) and outdoor air quality source; (3) data must be in 'raw' format and should be measured and documented as physical concentrations unit. Different air quality assessment indexes such as Air Quality Index (AQI), Pollutant Standard Index (PSI), or any other conventional measurement units (e.g. parts per million versus micrograms per cubic meter), which can easily convert to global standard physical measurement units was adopted by OpenAQ system; (4) the source and collection of data must be at the station-level, and the same should not be scaled up to higher (e.g. city) level to aggregate the measurements into a single value for easy assessment. The location-specific measurements allow to explore the intracity/location variation in pollution concentration; (5) the time-frequency of data collection should be between 10 minutes and 24 hours. The OpenAQ system checks the hosted sites in every 10 minutes to retrieve updated measurements of pollutants and 24 hours time frame considers the longest average time interval, which also defined as 'real-time' data. The geocoded air pollution estimates were converted to raster surface using Inverse distance weighted (IDW) interpolation. Three key air pollutants, i.e. carbon monoxide (CO), particulate matter ($PM_{2.5}$, $PM_{10}$), was considered to examine the effects of forest fire on air quality. All the data was collected for the period 15 August to 15 September for comparative assessment and subsequent interpretation.

---
[11] https://firms.modaps.eosdis.nasa.gov/active_fire/#firms-shapefile
[12] https://openaq.org/

*2.3 Measuring county scale wildfire hazard potential*

The county-scale wildfire hazard potential (WHP) scores were estimated using the WHP data developed by United States Department of Agriculture (USDA) Forest Service Fire Modeling Institute (FMI[13]) (Radeloff et al., 2017, 2018; Martinuzzi et al., 2015; Dillon et al., 2015). The FMI WHP data are providing a hazard potential index at a 270m spatial resolution for the entire conterminous USA. The WHP is also intended to measure the relative potential of wildfire hazard vulnerability at places that would often be difficult to contain. Therefore, FMI has stated that areas with high WHP scores exhibit higher risk and likelihood of impacting by high-intensity wildfires and vice versa. The final WHP scores is consisting of two WHP maps 1) continuous integer values, and 2) five WHP classes (very low = 1, low = 2, moderate = 3, high = 4, and very high = 5). The high (4) and very high WHP score (5) can be found in areas that frequently affected by wildfires, and 0 refers to the non-burnable areas or water surface. The low and moderate fire severity zones would be in 2 – 3 WHP score zone. According to USDA FMI, the main aim of developing a spatially explicit robust WHP score at high spatial resolution targeting the pixel as the smallest unit of measurement is not only describing the fire risk and fire behaviours of regions, but also assessing all other relevant components of fires risk by incorporating spatially paired demographic and resource information into the assessment. The improved multi-component WHP would therefore be able to be used for advance strategic planning and fuels management to curbs the concurrent problems of wildfire across the ecosystem (Radeloff et al., 2017, 2018; Martinuzzi et al., 2015; Dillon et al., 2015).

To pair the WHP scores with the demographic and other spatial components, the county scale demographic information provided by ESRI[14] was used. Linking up the population statistics into the wildfire hazard assessment allows us to spatially visualize and examine the risk level at each community that is prone to wildfires, especially at the west coastal states of USA (California, Washington, and Oregon). Additionally, having the improved status of wildfire hazard in places that encounter such problems for a long time would be beneficial for projecting cumulative loss that incurred due to wildfires and delineating avenues that could be adopted for managing forest fire at any given ecosystem (ESRI, 2020).

---

[13] https://www.firelab.org/project/wildfire-hazard-potential
[14] https://www.esri.com/arcgis-blog/products/arcgis-online/analytics/wildfire-hazard-potential-demographics/

To measure the county scale WHP, the average WHP values of each county was measured using zonal statistics as table function in ArcGISPro. The updated demographic profiles of each county provided by ESRI was incorporated into the modelling for computing the fire hazard score. The updated GIS layer consists of many geographic layers, including state, county, congressional district, ZIP code, tract, and block group was then joined with county-averaged WHP score tables for overlay analysis. The enrich tool function was applied for linking the updated county demographic attributes to county scale WHP scores and to measure the population-weighted average WHP scores at the county scale for better understanding the fire risk and associated adversities. The ESRI demographics include population count, household and housing units, population growth rate, and projected changes in population growth over time (ESRI, 2020).

## *2.4 Population weighted pollution exposure assessment*

The population-weighted air pollution exposure assessment has frequently used in epidemiological studies for measuring health effects of many extreme events (Gariazzo et al., 2016; Etchie et al., 2018; Sannigrahi et al., 2020a). In most cases, the exposure assessment was done using either relying on local monitoring measurements or satellite/in-situ data-driven Land Use Regression models (Gariazzo et al., 2016; Özkaynak et al., 2013). Through time, different modelling approaches have also been evolved to carry out air pollution exposure assessment at various scales. Researchers from Kings' College London used the data produced by CMAQ-urban model which later combined with KCLurban model for assessing human exposure assessment to $PM_{2.5}$, $PM_{10}$, $NO_2$, and NOx pollutants in London (Beevers et al., 2013). Two other commonly used models, i.e. AERMOD developed by US EPA and Common Noise Assessment Methods in EU member states (CNOSSOS-EU) (Kephalopoulos and Paviotti, 2012) have been used for traffic air pollution exposure assessment. In this study, the population-weighted average pollution concentration was measured for three key air pollutants, $PM_{2.5}$, $PM_{10}$, and CO. Both surface and satellite pollution estimates and gridded (pixel scale, 1km spatial resolution) raster population data provided by NASA SEDAC[15] was utilized to measure the grided population weighted pollution exposure during the fire period (15 August to 15 September 2019 and 2020). The county scale pollution exposure assessment was done using ArcGISPro spatial statistics zonal statistics as table function. The same was computed for different grid scale for examining the scale effects on population average pollution exposure. The updated administrative boundary was collected from US census bureau for measuring the county averaged pollution exposure for fire year (2020) and non-fire year (2019) for California, Oregon, and Washington states. The OECD (2016) provided algorithm was used to compute the population exposure assessment:

$$E_{POP} = \frac{\sum_x (Pop_x \times C_x)}{\sum_x Pop_x}$$

---

[15] https://sedac.ciesin.columbia.edu/data/set/gpw-v4-population-count-rev11

Where $E_{POP}$ is the population-weighted average concentration of PM$_{2.5}$, PM$_{10}$, NO$_2$, CO for each county of California, Oregon, and Washington. $x = 1…, N$ refers to the grids for which the exposure assessment was measured, $Pop_x$ is the population count derived from raster gridded population dataset from NASA SEDAC, $C_x$ is the average pollution concentration of PM$_{2.5}$, PM$_{10}$, NO$_2$, CO during the study period (15 August to 15 September 2019 and 2020). $\sum_x Pop_x$ is the total population count of the pixel/grid.

## *2.5 Experimental design*

The average concentration of the air pollutants was estimated using zonal statistics tool in ArcGIS Pro software. ArcPy python module was utilized for automation and easy calculation of the parameters utilized for comparative assessment. The spatiotemporal differences in air pollution concentration over the western coastal states, i.e. California, Oregon, and Washington, was measured using spatial analyst function in ArcGIS Pro. Spatial data visualization was done using ggplot2 package in R software and ArcGIS Pro 2.5.

## 3. Results and discussion

**Fig. 1** shows the black carbon density over the continental USA on 14, 15, and 16 September of 2020. Both storm progression path and plumes movements are demonstrating the strong linkages between meteorological components, i.e. wind speed, wind direction, etc. that facilitated the spread of smoke by sweeping air from west to east direction and plume progression over the time. Hurricane Paulette, which labelled as category 1 hurricane (wind speed would around 74 to 95 mph), developed in the Atlantic Ocean and made landfall in Bermuda in between the fire period, which also assisted to move high-flying smoke plumes from Midwest to Northeast as observed by Suomi NPP VIIRS satellite (NASA Earth Observatory, 2020). However, the occurrence of two catastrophes (forest fire and Hurricane Paulette) at the same period and the coupling effects and strong, cohesive interactions between the two events has blocked the progression of wildfire smoke towards the Atlantic Oceans in the upper atmosphere. Moreover, the Hurricane Sally, a category 2 hurricane (wind speed would be in between 96 and 110 mph), made landfall over the Gulf Coast and caused historic and catastrophic flooding the coastal region and helped the progression of smoke plumes towards the north side of the study region. Hurricanes Sally and Paulette have been the two of five tropical cyclones that occurred simultaneously in the Atlantic Ocean during the wildfire period and helped spread the wildfire plumes over the region.

The spatial concentration and differences in the key air pollutants, CO, PM$_{2.5}$, and PM$_{10}$ are presented in **Fig. 2**. Air pollution has been increasing substantially during the observation period, as evident for all three key air pollutants considered in this study. CO concentration over the fire-prone west coast region rose dramatically during the observation period (15 August to 15 September) in 2020, compared to the concentration recorded in 2019. The average CO concentration (ppm) in Washington, California, and Oregon was calculated as 2.34, 0.29, and 3.16 in 2019, which was increased up to 815.11, 13.46, and 1150.26 in 2020 (**Fig. 2, Table. 2**). The average $PM_{2.5}$ concentration ($\mu g/m^3$) over the same period was recorded as 4.29, 7.60, and 5.40 for Washington, California, and Oregon in 2019, which has risen up to 50.12, 38.87, and 47.39 in 2020 (**Fig. 2, Table. 2**). The concentration of $PM_{10}$ also shows a highly alarming state of air quality during the study period. The average $PM_{10}$ concentration ($\mu g/m^3$) was measured as 18.44, 29.40, and 23.28 in 2019, while the same has been increased considerably in 2020, recorded as 106.91, 64.44, and 86.03 for Washington, California, and Oregon (**Fig. 2, Table. 2**). Several earlier research has utilized many available resources, i.e. satellite estimates (Wu et al., 2006; Konovalov et al., 2011), in-situ measurement (Konovalov et al., 2011), low-cost sensor measurements (Sayahi et al., 2019; Delp and Singer, 2020), air pollution models (Watson et al., 2019) to examine the detrimental impact of wildfires on air quality across the ecosystems. Wu et al., (2006) study has noted that the concentration of $PM_{10}$ was increased up to 160 $\mu g/m^3$ due to the 2003 southern California wildfires and resulted emission of particulate matter. Hodzic et al., (2007) study on 2003 European wildfires documented a drastic increase (20 to 200%) of air pollutants, especially $PM_{10}$, due to the emission of gaseous compounds during the fire period. Wildfires were also found to be highly associated with the increases of fine particulate matter (d < 2.5µm) (Jaffe et al., 2008; Sullivan et al., 2008; Matz et al., 2020). Though there has been strong and clear evidence that wildfires have a strong negative impact on air quality, still, there are several other factors, such as description of fire emissions, atmospheric dispersion of smoke, and the chemical transformations of smoke, etc., needs to be evaluated comprehensively in order to understand the association between forest fire and air quality in a better way (Martins et al., 2012). The astonishing figures of these dramatic increases of air pollutants indicate that the record-breaking fires in 2020 not only caused massive destruction of wildlife and structural damages of properties, they're also adding enormous amounts of fire particulates including smoke and ash into the atmosphere, raising health crisis amidst the COVID-19 pandemic period. Though the clear connection between health effects and wildfire smoke has not fully explored, there are substantial shreds of evidence that supports the strong association between fire emitted smoke and ash concentration and severe health outcomes. It has also been observed in research that such effects have often exhibited delayed consequences, especially for cardiovascular and respiratory cause-specific diseases, and do not usually disappear when air gets clear (Landguth et al., 2020).

Satellite-based estimation also indicates the clustered concentration of air pollutants, i.e. AI, CO, and $NO_2$, during the fire incidents period (**Fig. 3, Table. 3**). The average Aerosol Index (AI) was recorded to be increased significantly for all three states, measured as -1.26, -0.94, and -1.18 in 2019 and -0.91, 0.44, and -0.53 in 2020 for Washington, California, and Oregon, respectively. The other two pollutants, i.e. CO and $NO_2$, are also found to be increased during the study period, which is mainly associated with wildfire and resulted in the emission of greenhouse compounds. The CO concentration ($\mu g/m^3$) was measured as 924.84, 855.46, 872.67 in 2019 and 1186.80, 1673.06, 1571.55 in 2020 for Washington, California, and

Oregon. While, the satellite-derived average $NO_2$ concentration (µg/m$^3$) was measured as 6.18, 6.29, and 5.86 in 2019 and 6.03, 6.83, and 5.97 in 2020 for Washington, California, and Oregon (**Fig. 3, Table. 3**). While, during the same period, the $O_3$ and $SO_2$ concentration was declined in 2020 for all three states. **Fig. 4** shows the carbon emission from fires in California state for the period 1997 to 2020. Carbon emission (Tg) was recorded highest in 2020, compared to the average carbon emissions during 1997 – 2020. This demonstrates the prolonged effects of forest fire on the atmospheric environment and GHG concentration that eventually led to mesoscale global warming across the ecosystems. (Li et al., 2020) study explored the application of advanced baseline imager (ABI) fire radiative power (FRP) for measuring wildfire emissions in the USA using S5P TROPOMI CO data and noted that during July 2018—October 2019, approximately 596 Gg CO has been added into the atmosphere due the wildfires (41 different sized wildfires haven been recorded during 2018-2019 fire season). During the fire period, Li et al. study measured the CO column mixing ratio of smoke plumes which was found 47 ppbv higher than that of the background (CO mixing ratio enhancement was 51% higher than the normal period). In extreme cases, the CO mixing ratio enhancement was measured as 164% (135 ppbv) and 261% (190 ppbv) for two fire events in Idaho region during the same period (Li et al., 2020). During the 2004-2009 fire period, on days when $PM_{2.5}$ AQI have exceeded the standard $PM_{2.5}$ AQI regulatory threshold, wildfires contributed nearly 71.3 % of total $PM_{2.5}$ AQI (Liu et al., 2016). Under the scenario of climate change, Liu et al. estimated that more than 82 million peoples would likely to experience a 57 % and 31 % rise in the frequency and intensity of Smoke waves. This unusual extremity will be more prominent in the fire-prone states of the USA, i.e. Northern California, Western Oregon and the Great Plains (Liu et al., 2016). Apart from that, many studies have utilized the COVID-19 lockdown restriction as a window to understanding the role of human activity on forest fire (Gupta et al., 2020; Amador-Jiménez et al., 2020).

The spatial differences in CO, $PM_{2.5}$, and $PM_{10}$ concentration were measured and documented in **Fig. 5** and **Fig. 6**. The in-situ data-led measurements show that the highest increases in CO (ppm), $PM_{2.5}$, and $PM_{10}$ concentration (µg/m$^3$) was clustered around the west coastal fire-prone states, from 15 August to 15 September period. The average CO concentration (ppm) was increased most significantly in Oregon (1147.10), followed by Washington (812.76), and California (13.17) (**Fig. 5, Table. 2**). While, the concentration (µg/m$^3$) of particulate matter (both $PM_{2.5}$ and $PM_{10}$), was increased in all three states that affected badly by wildfires. Changes (positive) in both $PM_{2.5}$ and $PM_{10}$ were measured highest in Washington (45.83 and 88.47 for $PM_{2.5}$ and $PM_{10}$), followed by Oregon (41.99 and 62.75 for $PM_{2.5}$ and $PM_{10}$), and California (31.27 and 35.04 for $PM_{2.5}$ and $PM_{10}$) (**Fig. 5, Table. 2**). The satellite-based assessment is also exhibiting to a high degree of changes in air pollution concentration observed over the west coastal states in the USA from 15 August to 15 September (**Fig. 6**). AI was increased mostly in California, compared to the changes in AI in Oregon and Washington in 2020 from the equivalent period in 2019. The changes in CO and $NO_2$ follows the same pattern as observed for AI. The highest increases in CO and $NO_2$ was recorded over the western coastal states, which is primarily associated with forest burns and release of GHG components into the atmosphere. The 7 days average AQI status during the fire period derived from PurpleAir has suggested the prolonged effects of wildfire smoke and blazes on ambient air quality (**Fig. 7**). Both $PM_{2.5}$ and $PM_{10}$ concentration over the west coastal states has reached the hazardous to most-hazardous AQI levels amidst the partial/full COVID-19 lockdown restrictions. The green dot in the map indicates that the air is mostly clean and no harm can be caused by this AQI. The orange color implies that few sensitive groups could be affected after 24 hours of exposure. The red color entails that everyone may be affected with this AQI if the exposure will be longer and there are no improvements in AQI. Among these colours, the dark purple-maroon indicates the worst AQI that suggest "Health warnings of emergency conditions if they are exposed for more than 24 hours and the entire population is more likely to be affected." (PurpleAir, 2020, CNBC, 2020). Due to wildfire and associated emission of particulate matter, a positive trend in $PM_{2.5}$ (98th quantile during 1988–2016) was observed in the Northwest United States (most fire-prone region in the USA), compared to the other areas of the country for which a negative trend in $PM_{2.5}$ was recorded (McClure and Jaffe, 2018). This positive trend in $PM_{2.5}$ concentration was found to be mainly associated with wildfire black carbon emission and emission of smoke in the west coast region (McClure and Jaffe, 2018). The wildfire smoke emission above the boundary layer often elevated into the free troposphere through convective lofting where the highspeed winds help to travel the smoke to a long distance. The subsidence of this smoke layer occurs at the zone of high surface pressure, which can contribute to the enhancement of surface concentration of fine particulate matter (Miller et al., 2011).

Population weighted average concentration (ppm for CO and $\mu g/m^3$ for $PM_{2.5}$ and $PM_{10}$) of air pollutants was measured for California, Washington, and Oregon for 15 August to 15 September period (**Table. 4**). The exposure assessment has been carried out for two consecutive years, i.e. 2019 (considered as the normal year) and 2020 (considered as fire year). The average exposure level of CO has been increased dramatically in Washington and Oregon, while, the same has not been increased greatly in California. The highest average level of exposure to $PM_{2.5}$ was found in Oregon, followed by Washington and California. At the same time, the highest exposure level of $PM_{10}$ exposure was calculated for Washington, Oregon, and California, states respectively (**Table. 4**). According to WHO classification of global air pollution regulatory threshold for $PM_{2.5}$, i.e. (i) highly polluted – (average $PM_{2.5} > 35$ $\mu g/m^3$); (ii) polluted ($35 \geq$ average $PM_{2.5} > 25$ $\mu g/m^3$); (iii) moderately polluted ($25 \geq$ average $PM_{2.5} > 15$ $\mu g/m^3$); and (iv) slightly polluted ($15 \geq$ average $PM_{2.5} > 10$ $\mu g/m^3$), air quality status in Oregon and Washington have reached the highly polluted condition during the fire period. While $PM_{2.5}$ AQI was measured as polluted (average $PM_{2.5}$ concentration in California was recorded as 28.90 $\mu g/m^3$ during the study period in 2020). In 2019, during the same period, the average exposure ($\mu g/m^3$) to $PM_{2.5}$ was documented as 4.77, 9.06, and 5.04 for Washington, California, and Oregon, respectively (**Table. 4**). According to the US EPA AQI categorization for $PM_{2.5}$ ($\mu g/m^3$), i.e. good ($PM_{2.5}$ – 0-12), moderate ($PM_{2.5}$ – 12 – 35), unhealthy for sensitive individuals ($PM_{2.5}$ – 35 – 55), unhealthy ($PM_{2.5}$ – 55-150), very unhealthy ($PM_{2.5}$ – 150 – 250), and hazardous ($PM_{2.5}$ – >250), Oregon's ($PM_{2.5} = 52.14$) and Washington's ($PM_{2.5} = 41.02$) $PM_{2.5}$ AQI was categorized as unhealthy for sensitive individuals (**Table. 4**). According to the WHO air quality guidelines for $PM_{10}$ ($PM_{10}$ should not be exceeded 20 $\mu g/m^3$ annual mean, or 50 $\mu g/m^3$ 24-hour mean), the average exposure ($\mu g/m^3$) level to $PM_{10}$ was found unhealthy and hazardous for all three states, i.e. Washington (102.77), Oregon (93.32), and California (68.04), respectively during 15 August to 15 September in 2020 (**Table. 4**). This sudden and abrupt increases of fine and coarse particulate matter and carbon compounds is largely due to the wildfire emission, and this unprecedented event poses serious health concerns to the residents living in this region. Currently, wildfire smoke causes 339,000 annual global premature mortality (interquartile range: 260,000–600,000) (Johnston et al., 2012). Liu et al. (2015) evaluated 61 epidemiological studies linked with wildfire and human health across the world, and reported that daily air pollution levels that recorded during or after wildfire events exceeded U.S. EPA regulations and in many cases, the average $PM_{10}$ concentration during the wildfire period was found 1.2 to 10 times higher than that of non-fire periods. Among the diseases that primarily attributed to wildfire, respiratory disease was found to be highly associated with wildfire smoke (Liu et al., 2015). In the USA, nearly 10% of the population (30.5 million) stayed in the regions where the contribution of wildfire to annual average ambient $PM_{2.5}$ was high (>1.5 μg/m$^3$) and nearly 10.3 million peoples experienced to unhealthy air quality for more than 10 consecutive days that mainly attributed to wildfire and resulted in emission (Rappold et al., 2017). People with existing respiratory illness have found more susceptible due to wildfire, and the upsurges of the demand of rescue medication have been linked to the average exposure time to wildfire smoke in Southern California, USA (Vora et al., 2011). Ischemic Heart Disease (IHD) had also been linked to the wildfire emission (particularly $PM_{10}$ emission) as the elevated number of IHD related clinic visits reported during the wildfire seasons in California (Lee et al., 2009).

## Conclusion

The present research has evaluated the effects of forest fire on air quality status in the west coastal states (California, Oregon, and Washington) of the USA. 2020 west coast fire has broken the past fire records, and nearly 6 million forest areas were burned in California, Oregon, and Washington states in the USA. To understand the adverse effects of wildfire emissions on air quality, the concentrations of different pollutants, i.e. CO, $PM_{2.5}$, $PM_{10}$, $NO_2$, $O_3$, $SO_2$, and AI, were measured using satellite and in-situ monitored data for a fixed period (15 August to 15 September) for 2020 (fire year) and 2019 (reference year). Sentinel 5P TROPOMI satellite data products were utilized for estimating column number density of $NO_2$, $O_3$, $SO_2$, CO, and AI. The average CO concentration (ppm) was increased most notably in Oregon (1147.10), followed by Washington (812.76), and California (13.17). While, the concentration (μg/m$^3$) of particulate matter (both $PM_{2.5}$ and $PM_{10}$), was increased in all three states that affected severely by wildfires. Changes (positive) in both $PM_{2.5}$ and $PM_{10}$ were measured highest in Washington (45.83 and 88.47 for $PM_{2.5}$ and $PM_{10}$), followed by Oregon (41.99 and 62.75 for $PM_{2.5}$ and $PM_{10}$), and California (31.27 and 35.04 for $PM_{2.5}$ and $PM_{10}$), respectively. The county scale hazard potential assessment has portrayed the abundance of dry fuel load and potential risk of wildland fires across the region. Due to the unprecedented wildland fire and resulted emission, the average level of exposure to hazardous pollutants have been increased drastically. This may cause severe health hazards if the fire smoke persists a longer period. The high-intensity fire events in west coast regions have been triggered by many causal factors, including heatwave and warming climate prevalent in the region, the abundant load of dry fuel and lack of soil moisture, cold breezes from nearby states and occurrence of hurricanes in the midst of fire events, etc. The increasing level of air pollution, caused by the west coast fire in 2020, revealed the strong negative association between wildfire and air quality status of a region. More domain research in this direction would help to understand the complex nexus between climate change and associated adversities in a better way.

**References:**

Aditama, T.Y., 2000. Impact of haze from forest fire to respiratory health: Indonesian experience. Respirology 5, 169–174. https://doi.org/10.1046/j.1440-1843.2000.00246.x

Amador-Jiménez, M., Millner, N., Palmer, C., Pennington, R.T., Sileci, L., 2020. The Unintended Impact of Colombia's Covid-19 Lockdown on Forest Fires. Environ. Resour. Econ. 76, 1081–1105. https://doi.org/10.1007/s10640-020-00501-5

British Broadcasting Corporation (BBC), 2020. https://www.bbc.com/news/world-us-canada-54180049

B., H.S., Michael, B., C., M.Y., M., K.S., 2011. Three Measures of Forest Fire Smoke Exposure and Their Associations with Respiratory and Cardiovascular Health Outcomes in a Population-Based Cohort. Environ. Health Perspect. 119, 1266–1271. https://doi.org/10.1289/ehp.1002288

Beevers, S.D., Kitwiroon, N., Williams, M.L., Kelly, F.J., Ross Anderson, H., Carslaw, D.C., 2013. Air pollution dispersion models for human exposure predictions in London. J. Expo. Sci. Environ. Epidemiol. 23, 647–653. https://doi.org/10.1038/jes.2013.6

Bo, M., Mercalli, L., Pognant, F., Cat Berro, D., Clerico, M., 2020. Urban air pollution, climate change and wildfires: The case study of an extended forest fire episode in northern Italy favoured by drought and warm weather conditions. Energy Reports 6, 781–786. https://doi.org/https://doi.org/10.1016/j.egyr.2019.11.002

Butt, E.W., Conibear, L., Reddington, C.L., Darbyshire, E., Morgan, W.T., Coe, H., Artaxo, P., Brito, J., Knote, C., Spracklen, D. V, 2020. Large air quality and human health impacts due to Amazon forest and vegetation fires. Environ. Res. Commun. 2, 95001. https://doi.org/10.1088/2515-7620/abb0db

Cable News Network (CNN), 2020. https://edition.cnn.com/2020/09/18/us/west-coast-wildfires-friday/index.html

California Department of Forestry and Fire Protection (CAL FIRE), 2020. https://www.fire.ca.gov/incidents/

Cascio, W.E., 2018. Wildland fire smoke and human health. Sci. Total Environ. 624, 586–595. https://doi.org/https://doi.org/10.1016/j.scitotenv.2017.12.086

Cheng, L., McDonald, K.M., Angle, R.P., Sandhu, H.S., 1998. Forest fire enhanced photochemical air pollution. A case study. Atmos. Environ. 32, 673–681. https://doi.org/https://doi.org/10.1016/S1352-2310(97)00319-1

DeBell, L.J., Talbot, R.W., Dibb, J.E., Munger, J.W., Fischer, E. V, Frolking, S.E., 2004. A major regional air pollution event in the northeastern United States caused by extensive forest fires in Quebec, Canada. J. Geophys. Res. Atmos. 109.

https://doi.org/10.1029/2004JD004840

Delp, W.W., Singer, B.C., 2020. Professional PM 2 . 5 Monitors with Optical Sensors.

Dillon, G.K., Menakis, J., Fay, F., 2015. Wildland Fire Potential : A Tool for Assessing Wildfire Risk and Fuels Management Needs. Proc. large Wildl. fires Conf. 60–76.

Dillon, G.K.; J. Menakis; and F. Fay. 2015. Wildland Fire Potential: A Tool for Assessing Wildfire Risk and Fuels Management Needs. pp 60-76

In Keane, R. E.; Jolly, M.; Parsons, R.; and Riley, K. Proceedings of the large wildland fires conference; May 19-23, 2014; Missoula, MT. Proc. RMRS-P-73. Fort Collins, CO: U.S. Department of Agriculture, Forest Service, Rocky Mountain Research Station. 345 p.

Etchie, T.O., Etchie, A.T., Adewuyi, G.O., Pillarisetti, A., Sivanesan, S., Krishnamurthi, K., Arora, N.K., 2018. The gains in life expectancy by ambient PM2.5 pollution reductions in localities in Nigeria. Environ. Pollut. 236, 146–157. https://doi.org/https://doi.org/10.1016/j.envpol.2018.01.034

European Comission Atmospheric Monitoring Services, 2020. https://atmosphere.copernicus.eu/

EPA, 2020. https://www.epa.gov/

Ferrare, R.A., Fraser, R.S., Kaufman, Y.J., 1990. Satellite measurements of large-scale air pollution: Measurements of forest fire smoke. J. Geophys. Res. Atmos. 95, 9911–9925. https://doi.org/10.1029/JD095iD07p09911

Fowler, C., 2003. Human Health Impacts of Forest Fires in the Southern United States: A Literature Review. J. Ecol. Anthropol. 7, 39–63. https://doi.org/10.5038/2162-4593.7.1.3

Guanter L, Aben I, Tol P, Krijger JM, Hollstein A, Köhler P, Damm A, Joiner J, Frankenberg C, Landgraf J. 2015. Potential of the TROPOspheric Monitoring Instrument (TROPOMI) onboard the Sentinel-5 Precursor for the monitoring of terrestrial chlorophyll fluorescence. Atmos Meas Tech. 8(3):1337–1352. doi:10.5194/amt-8-1337-2015.

Gariazzo, C., Pelliccioni, A., Bolignano, A., 2016. A dynamic urban air pollution population exposure assessment study using model and population density data derived by mobile phone traffic. Atmos. Environ. 131, 289–300. https://doi.org/https://doi.org/10.1016/j.atmosenv.2016.02.011

Gorelick, N., Hancher, M., Dixon, M., Ilyushchenko, S., Thau, D., & Moore, R. (2017). Google Earth Engine: Planetary-scale geospatial analysis for everyone. Remote sensing of Environment, 202, 18-27.

Gupta, A., Bhatt, C.M., Roy, A., Chauhan, P., 2020. COVID-19 lockdown a window of opportunity to understand the role of human activity on forest fire incidences in the Western Himalaya, India. Curr. Sci. 119, 390–398.

Hodzic, A., Madronich, S., Bohn, B., Massie, S., Menut, L., Wiedinmyer, C., 2007. Wildfire particulate matter in Europe during summer 2003: meso-scale modeling of smoke emissions, transport and radiative effects. Atmos. Chem. Phys. 7, 4043–4064. https://doi.org/10.5194/acp-7-4043-2007

Jaffe, D., Hafner, W., Chand, D., Westerling, A., Spracklen, D., 2008. Interannual variations in PM2.5 due to wildfires in the Western United States. Environ. Sci. Technol. 42, 2812–

2818. https://doi.org/10.1021/es702755v

Johnston, F.H., Henderson, S.B., Chen, Y., Randerson, J.T., Marlier, M., DeFries, R.S., Kinney, P., Bowman, D.M.J.S., Brauer, M., 2012. Estimated Global Mortality Attributable to Smoke from Landscape Fires. Environ. Health Perspect. 120, 695–701. https://doi.org/10.1289/ehp.1104422

Johnston, F.H., Purdie, S., Jalaludin, B., Martin, K.L., Henderson, S.B., Morgan, G.G., 2014. Air pollution events from forest fires and emergency department attendances in Sydney, Australia 1996–2007: a case-crossover analysis. Environ. Heal. 13, 105. https://doi.org/10.1186/1476-069X-13-105

Kaplan G, Avdan ZY, Avdan U. 2019. Spaceborne Nitrogen Dioxide Observations from the Sentinel-5P TROPOMI over Turkey. Proceedings. 18(1):4. doi:10.3390/ecrs-3-06181.

Kephalopoulos, S., Paviotti, M., 2012. Noise Assessment. https://doi.org/10.2788/31776

Konovalov, I.B., Beekmann, M., Kuznetsova, I.N., Yurova, A., Zvyagintsev, A.M., 2011. Atmospheric impacts of the 2010 Russian wildfires: integrating modelling and measurements of an extreme air pollution episode in the Moscow region. Atmos. Chem. Phys. 11, 10031–10056. https://doi.org/10.5194/acp-11-10031-2011

Kumar, P., Hama, S., Omidvarborna, H., Sharma, A., Sahani, J., Abhijith, K.V., Debele, S., Zavala-Reyes, J., Barwise, Y., Tiwari, A., 2020. Temporary reduction in fine particulate matter due to 'anthropogenic emissions switch-off' during COVID-19 lockdown in Indian cities. Sustain. Cities Soc. 62, 102382. https://doi.org/10.1016/j.scs.2020.102382

Landguth, E.L., Holden, Z.A., Graham, J., Stark, B., Mokhtari, E.B., Kaleczyc, E., Anderson, S., Urbanski, S., Jolly, M., Semmens, E.O., Warren, D.A., Swanson, A., Stone, E., Noonan, C., 2020. The delayed effect of wildfire season particulate matter on subsequent influenza season in a mountain west region of the USA. Environ. Int. 139, 105668. https://doi.org/https://doi.org/10.1016/j.envint.2020.105668

Lazaridis, M., Latos, M., Aleksandropoulou, V., Hov, Ø., Papayannis, A., Tørseth, K., 2008. Contribution of forest fire emissions to atmospheric pollution in Greece. Air Qual. Atmos. Heal. 1, 143–158. https://doi.org/10.1007/s11869-008-0020-0

Lee, T.-S., Falter, K., Meyer, P., Mott, J., Gwynn, C., 2009. Risk factors associated with clinic visits during the 1999 forest fires near the Hoopa Valley Indian Reservation, California, USA. Int. J. Environ. Health Res. 19, 315–327. https://doi.org/10.1080/09603120802712750

Li, F., Zhang, X., Kondragunta, S., Lu, X., 2020. An evaluation of advanced baseline imager fire radiative power based wildfire emissions using carbon monoxide observed by the Tropospheric Monitoring Instrument across the conterminous United States. Environ. Res. Lett. 15, 94049. https://doi.org/10.1088/1748-9326/ab9d3a

Li, L.-M., Song, W.-G., Ma, J., Satoh, K., 2009. Artificial neural network approach for modeling the impact of population density and weather parameters on forest fire risk. Int. J. Wildl. Fire 18, 640–647.

Liu, J.C., Mickley, L.J., Sulprizio, M.P., Dominici, F., Yue, X., Ebisu, K., Anderson, G.B., Khan, R.F.A., Bravo, M.A., Bell, M.L., 2016. Particulate air pollution from wildfires in the Western US under climate change. Clim. Change 138, 655–666. https://doi.org/10.1007/s10584-016-1762-6

Liu, J.C., Pereira, G., Uhl, S.A., Bravo, M.A., Bell, M.L., 2015. A systematic review of the physical health impacts from non-occupational exposure to wildfire smoke. Environ. Res. 136, 120–132. https://doi.org/https://doi.org/10.1016/j.envres.2014.10.015

Loehman, R.A., Clark, J.A., Keane, R.E., 2011. Modeling effects of climate change and fire management on western white pine (Pinus monticola) in the Northern Rocky Mountains, USA. Forests 2, 832–860. https://doi.org/10.3390/f2040832

Martinuzzi, Sebastián; Stewart, Susan I.; Helmers, David P.; Mockrin, Miranda H.; Hammer, Roger B.; Radeloff, Volker C. 2015. The 2010 wildland-urban interface of the conterminous United States. Research Map NRS-8. Newtown Square, PA: U.S. Department of Agriculture, Forest Service, Northern Research Station. 124 p. (https://www.nrs.fs.fed.us/data/WUI/).

Martins, V., Miranda, A.I., Carvalho, A., Schaap, M., Borrego, C., Sá, E., 2012. Impact of forest fires on particulate matter and ozone levels during the 2003, 2004 and 2005 fire seasons in portugal. Sci. Total Environ. 414, 53–62. https://doi.org/10.1016/j.scitotenv.2011.10.007

Matz, C.J., Egyed, M., Xi, G., Racine, J., Pavlovic, R., Rittmaster, R., Henderson, S.B., Stieb, D.M., 2020. Health impact analysis of PM2.5 from wildfire smoke in Canada (2013–2015, 2017–2018). Sci. Total Environ. 725, 138506. https://doi.org/https://doi.org/10.1016/j.scitotenv.2020.138506

McClure, C.D., Jaffe, D.A., 2018. US particulate matter air quality improves except in wildfire-prone areas. Proc. Natl. Acad. Sci. 115, 7901 LP – 7906. https://doi.org/10.1073/pnas.1804353115

Miller, D.J., Sun, K., Zondlo, M.A., Kanter, D., Dubovik, O., Welton, E.J., Winker, D.M., Ginoux, P., 2011. Assessing boreal forest fire smoke aerosol impacts on U.S. air quality: A case study using multiple data sets. J. Geophys. Res. Atmos. 116. https://doi.org/10.1029/2011JD016170

National Aeronautics and Space Administration Earth Observatory, 2020. https://earthobservatory.nasa.gov/images/147277/historic-fires-devastate-the-us-pacific-coast

OpenAQ, 2020. https://openaq.org/#/countries?_k=ok3o24

Özkaynak, H., Baxter, L.K., Dionisio, K.L., Burke, J., 2013. Air pollution exposure prediction approaches used in air pollution epidemiology studies. J. Expo. Sci. Environ. Epidemiol. 23, 566–572. https://doi.org/10.1038/jes.2013.15

Phuleria, H.C., Fine, P.M., Zhu, Y., Sioutas, C., 2005. Air quality impacts of the October 2003 Southern California wildfires. J. Geophys. Res. Atmos. 110. https://doi.org/10.1029/2004JD004626

PurpleAir, 2020. https://www2.purpleair.com/

Radeloff, Volker C.; Helmers, David P.; Kramer, H. Anu; Mockrin, Miranda H.; Alexandre, Patricia M.; Bar-Massada, Avi; Butsic, Van; Hawbaker, Todd J.; Martinuzzi, Sebastián; Syphard, Alexandra D.; Stewart, Susan I. 2018. Rapid growth of the US wildland-urban interface raises wildfire risk. Proceedings of the National Academy of Sciences. 24: 201718850-. https://doi.org/10.1073/pnas.1718850115.

Radeloff, Volker C.; Helmers, David P.; Kramer, H. Anu; Mockrin, Miranda H.; Alexandre,

Patricia M.; Bar Massada, Avi; Butsic, Van; Hawbaker, Todd J.; Martinuzzi, Sebastián; Syphard, Alexandra D.; Stewart, Susan I. 2017. The 1990-2010 wildland-urban interface of the conterminous United States - geospatial data. 2nd Edition. Fort Collins, CO: Forest Service Research Data Archive. https://doi.org/10.2737/RDS-2015-0012-2.

Rappold, A.G., Reyes, J., Pouliot, G., Cascio, W.E., Diaz-Sanchez, D., 2017. Community Vulnerability to Health Impacts of Wildland Fire Smoke Exposure. Environ. Sci. Technol. 51, 6674–6682. https://doi.org/10.1021/acs.est.6b06200

Sannigrahi, S., Molter, A., Kumar, P., Zhang, Q., Basu, B., Sarkar Basu, A., Pilla, F., 2020a. Examining the status of improved air quality due to COVID-19 lockdown and an associated reduction in anthropogenic emissions. medRxiv 2020.08.20.20177949.

Sannigrahi, S., Pilla, F., Basu, B., Basu, A.S., Sarkar, K., Chakraborti, S., Joshi, P.K., Zhang, Q., Wang, Y., Bhatt, S., Bhatt, A., Jha, S., Keesstra, S., Roy, P.S., 2020b. Examining the effects of forest fire on terrestrial carbon emission and ecosystem production in India using remote sensing approaches. Sci. Total Environ. 725, 138331. https://doi.org/https://doi.org/10.1016/j.scitotenv.2020.138331

Sapkota, A., Symons, J.M., Kleissl, J., Wang, L., Parlange, M.B., Ondov, J., Breysse, P.N., Diette, G.B., Eggleston, P.A., Buckley, T.J., 2005. Impact of the 2002 Canadian Forest Fires on Particulate Matter Air Quality in Baltimore City. Environ. Sci. Technol. 39, 24–32. https://doi.org/10.1021/es035311z

Sayahi, T., Butterfield, A., Kelly, K.E., 2019. Long-term field evaluation of the Plantower PMS low-cost particulate matter sensors. Environ. Pollut. 245, 932–940. https://doi.org/https://doi.org/10.1016/j.envpol.2018.11.065

Schweizer, D.W., Cisneros, R., 2017. Forest fire policy: change conventional thinking of smoke management to prioritize long-term air quality and public health. Air Qual. Atmos. Heal. 10, 33–36. https://doi.org/10.1007/s11869-016-0405-4

Sommers, W.T., Loehman, R.A., Hardy, C.C., 2014. Wildland fire emissions, carbon, and climate: Science overview and knowledge needs. For. Ecol. Manage. 317, 1–8. https://doi.org/10.1016/j.foreco.2013.12.014

Sullivan, A.P., Holden, A.S., Patterson, L.A., McMeeking, G.R., Kreidenweis, S.M., Malm, W.C., Hao, W.M., Wold, C.E., Collett Jr., J.L., 2008. A method for smoke marker measurements and its potential application for determining the contribution of biomass burning from wildfires and prescribed fires to ambient PM2.5 organic carbon. J. Geophys. Res. Atmos. 113. https://doi.org/10.1029/2008JD010216

Urbanski, S.P., Hao, W.M., Baker, S., 2008. Chapter 4 Chemical Composition of Wildland Fire Emissions, in: Bytnerowicz, A., Arbaugh, M.J., Riebau, A.R., Andersen, C.B.T.-D. in E.S. (Eds.), Wildland Fires and Air Pollution. Elsevier, pp. 79–107. https://doi.org/https://doi.org/10.1016/S1474-8177(08)00004-1

Verma, S., Jayakumar, S., 2012. Impact of forest fire on physical, chemical and biological properties of soil: A. Proc. Int. Acad. … 2, 168–176.

Vora, C., Renvall, M.J., Chao, P., Ferguson, P., Ramsdell, J.W., 2011. 2007 San Diego Wildfires and Asthmatics. J. Asthma 48, 75–78. https://doi.org/10.3109/02770903.2010.535885

Watson, G.L., Telesca, D., Reid, C.E., Pfister, G.G., Jerrett, M., 2019. Machine learning

models accurately predict ozone exposure during wildfire events. Environ. Pollut. 254, 112792. https://doi.org/https://doi.org/10.1016/j.envpol.2019.06.088

Wu, J., M Winer, A., J Delfino, R., 2006. Exposure assessment of particulate matter air pollution before, during, and after the 2003 Southern California wildfires. Atmos. Environ. 40, 3333–3348. https://doi.org/https://doi.org/10.1016/j.atmosenv.2006.01.056

Zhu, Q., Liu, Y., Jia, R., Hua, S., Shao, T., Wang, B., 2018. A numerical simulation study on the impact of smoke aerosols from Russian forest fires on the air pollution over Asia. Atmos. Environ. 182, 263–274. https://doi.org/https://doi.org/10.1016/j.atmosenv.2018.03.052

https://doc.arcgis.com/en/esri-demographics/data/us-intro.htm

https://doc.arcgis.com/en/arcgis-online/analyze/enrich-layer.htm

https://www.esri.com/arcgis-blog/products/arcgis-online/analytics/wildfire-hazard-potential-demographics/

https://www.firelab.org/project/wildfire-hazard-potential

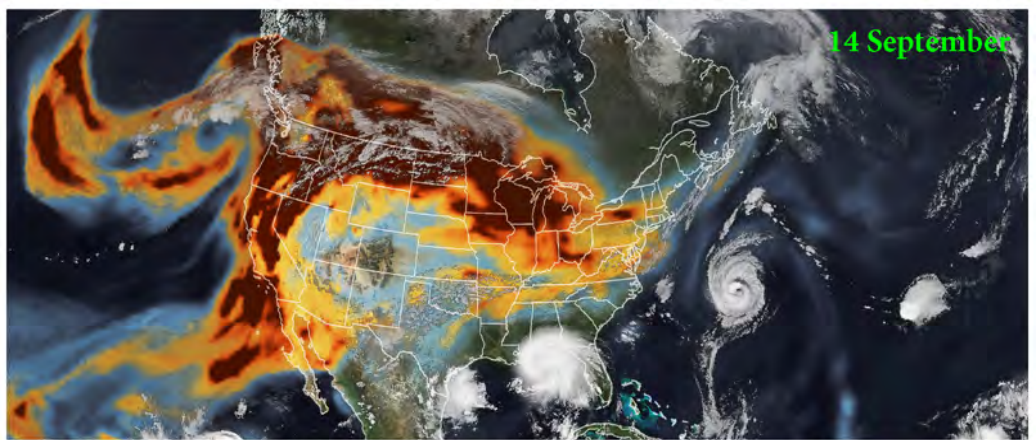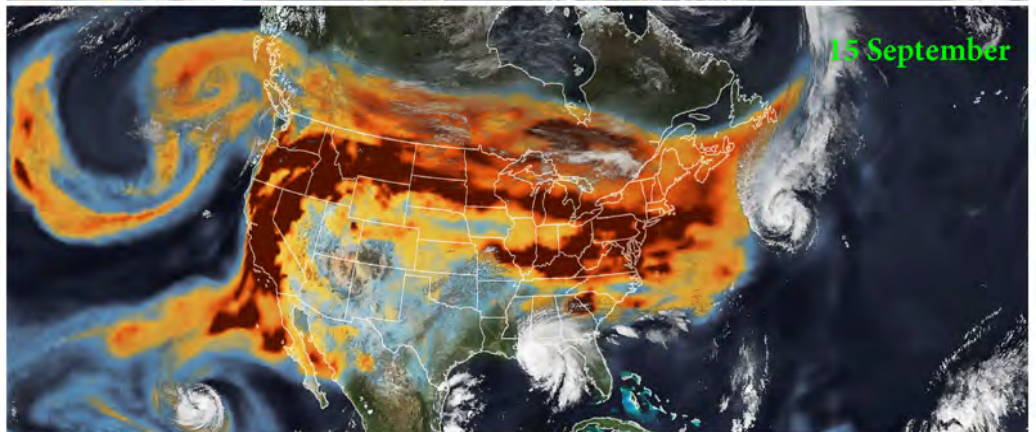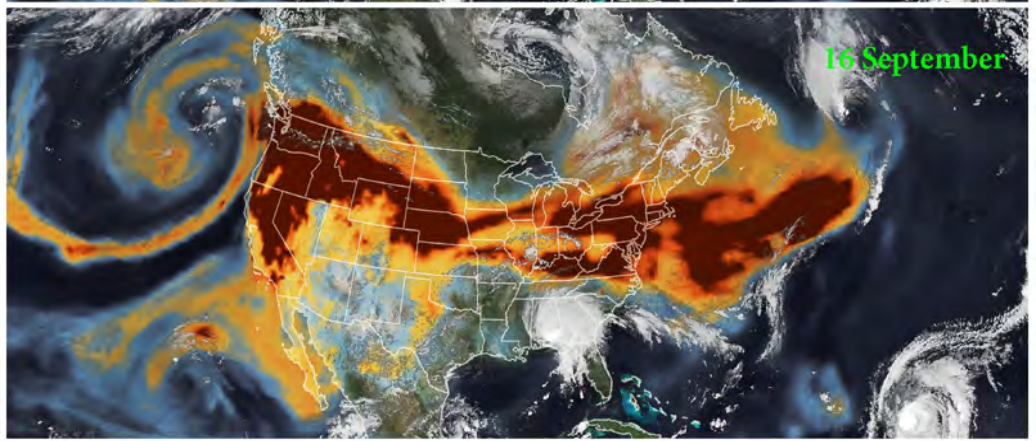

**Fig. 1** Spatial distribution of black carbon over the country, derived from GEOS forward processing (GEOS-FP) model and storm images derived from Suomi NPP — VIIRS satellite data. **Source** - NASA, 2020.

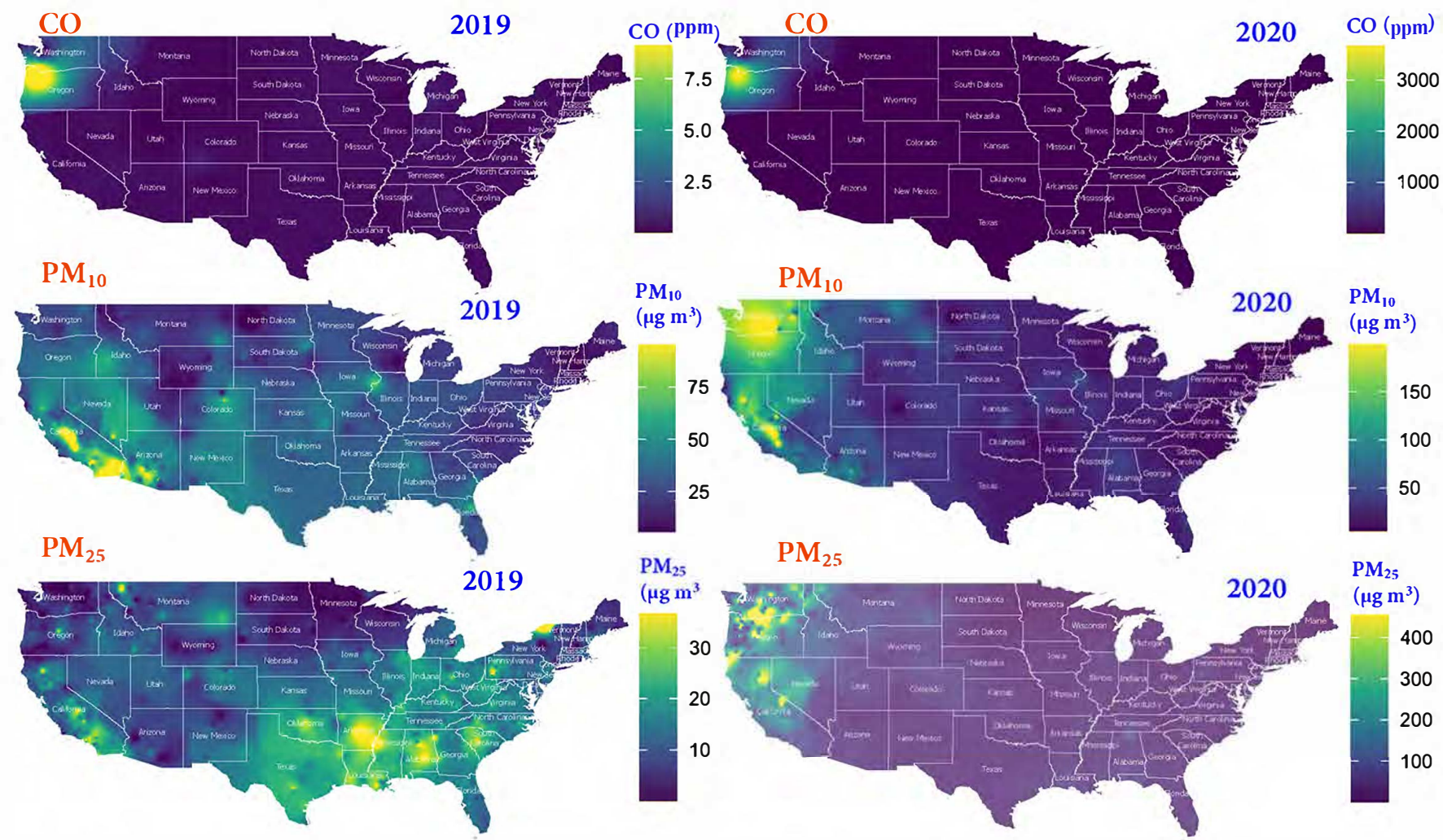

Fig. 2 Spatial differences in pollution concentration during 15 August to 15 September 2019 and 2020.

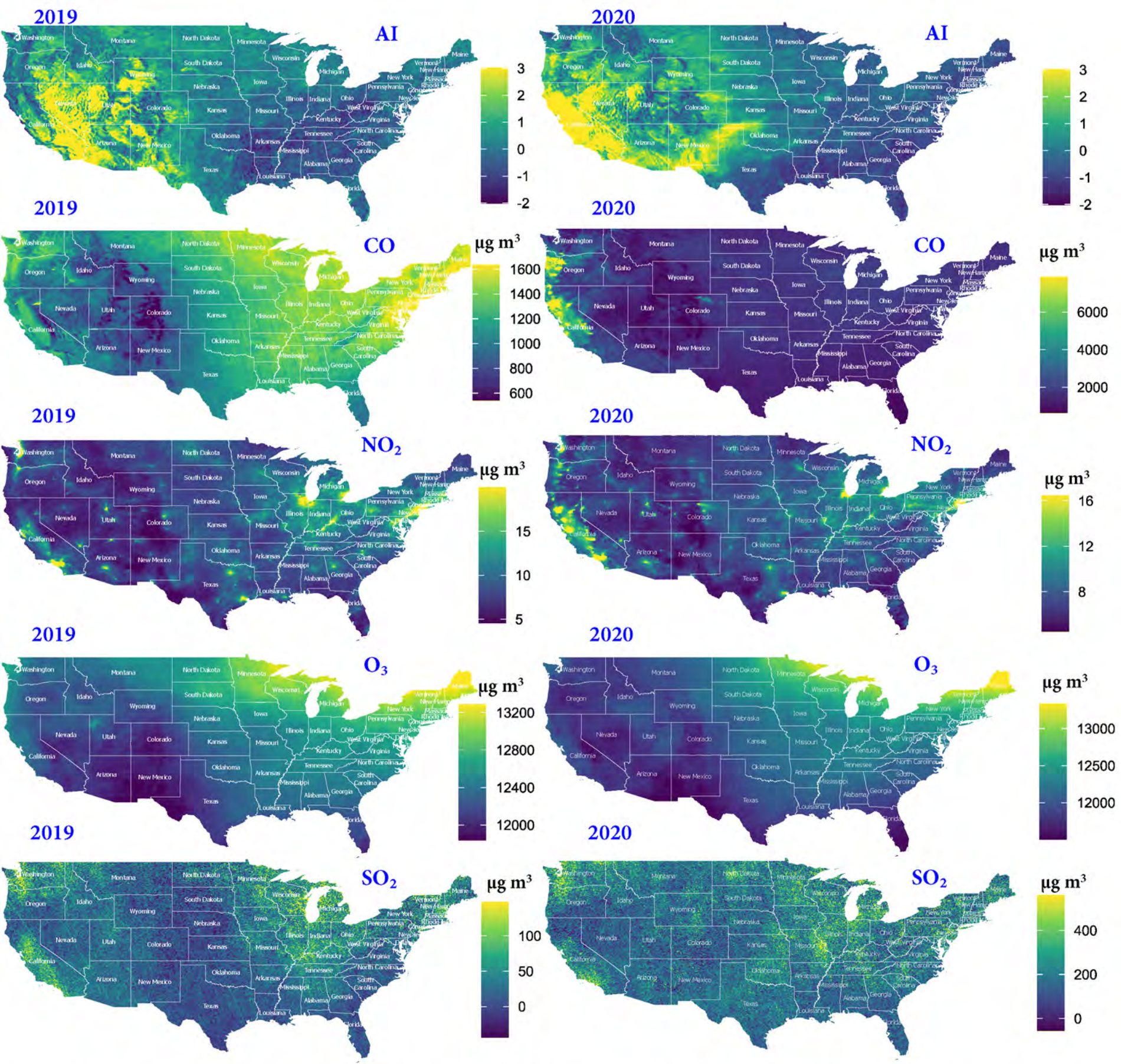

Fig. 3 Pollution concentration during the study period (15 August to 15 September) derived from Sentinel 5P data.

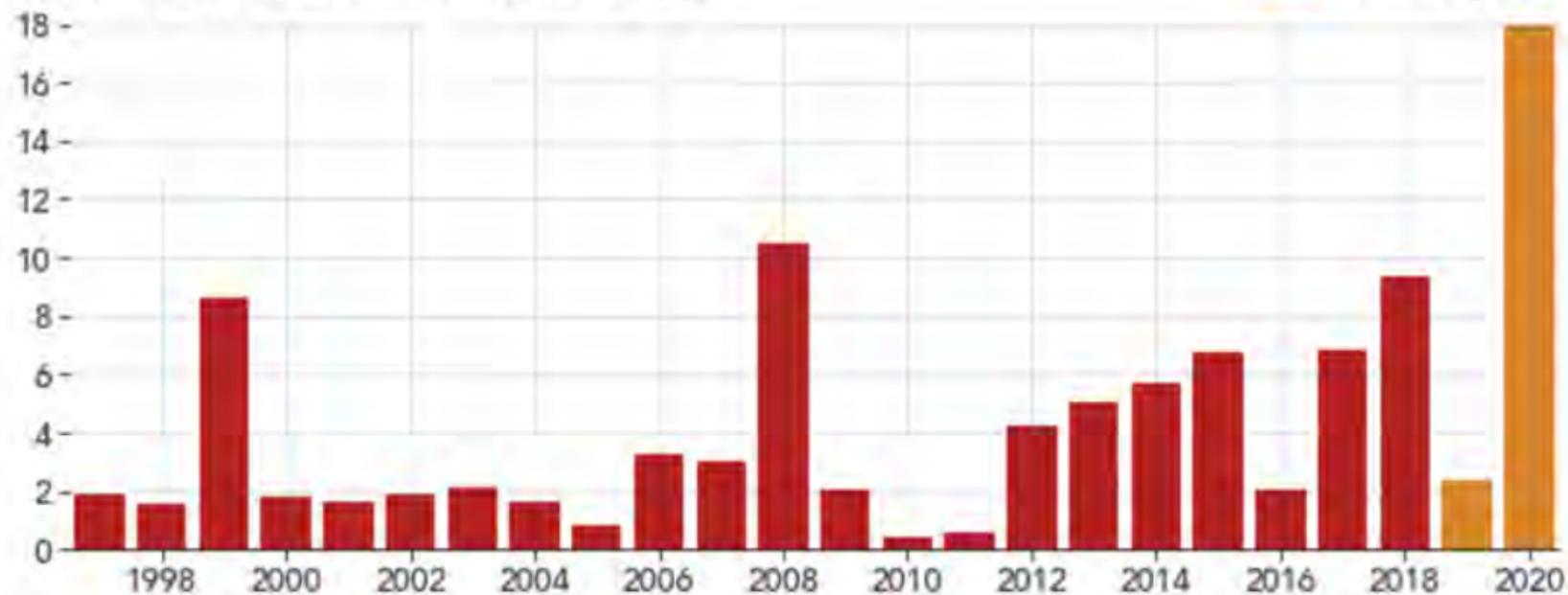

**Fig. 4** Carbon emissions statistics in California during 1997 to 2020. Carbon emission due to west coast fire was found highest during the last 20 years. Source - NASA, 2020.

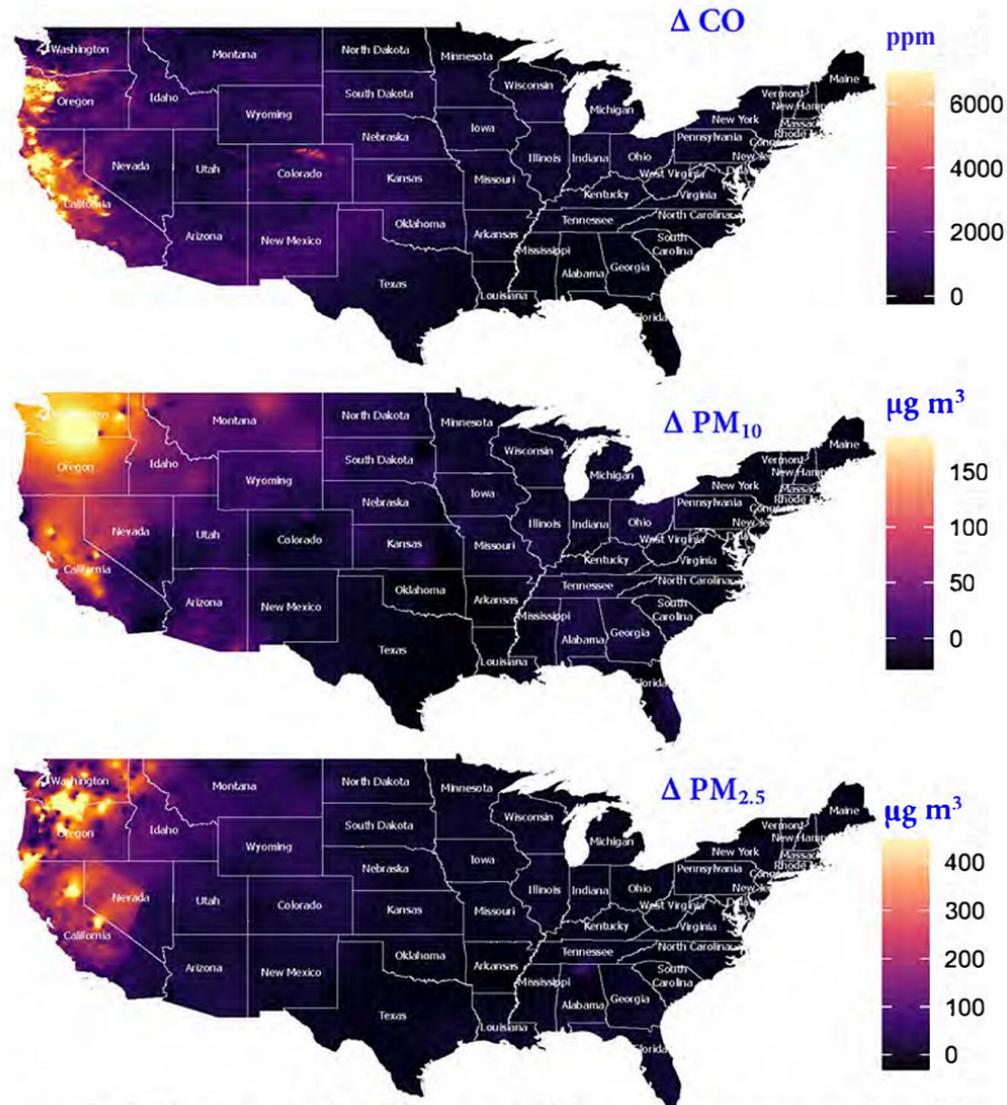

Fig. 5 Differences in CO, PM$_{10}$, and PM$_{2.5}$ concentration due to west coast forest fire.

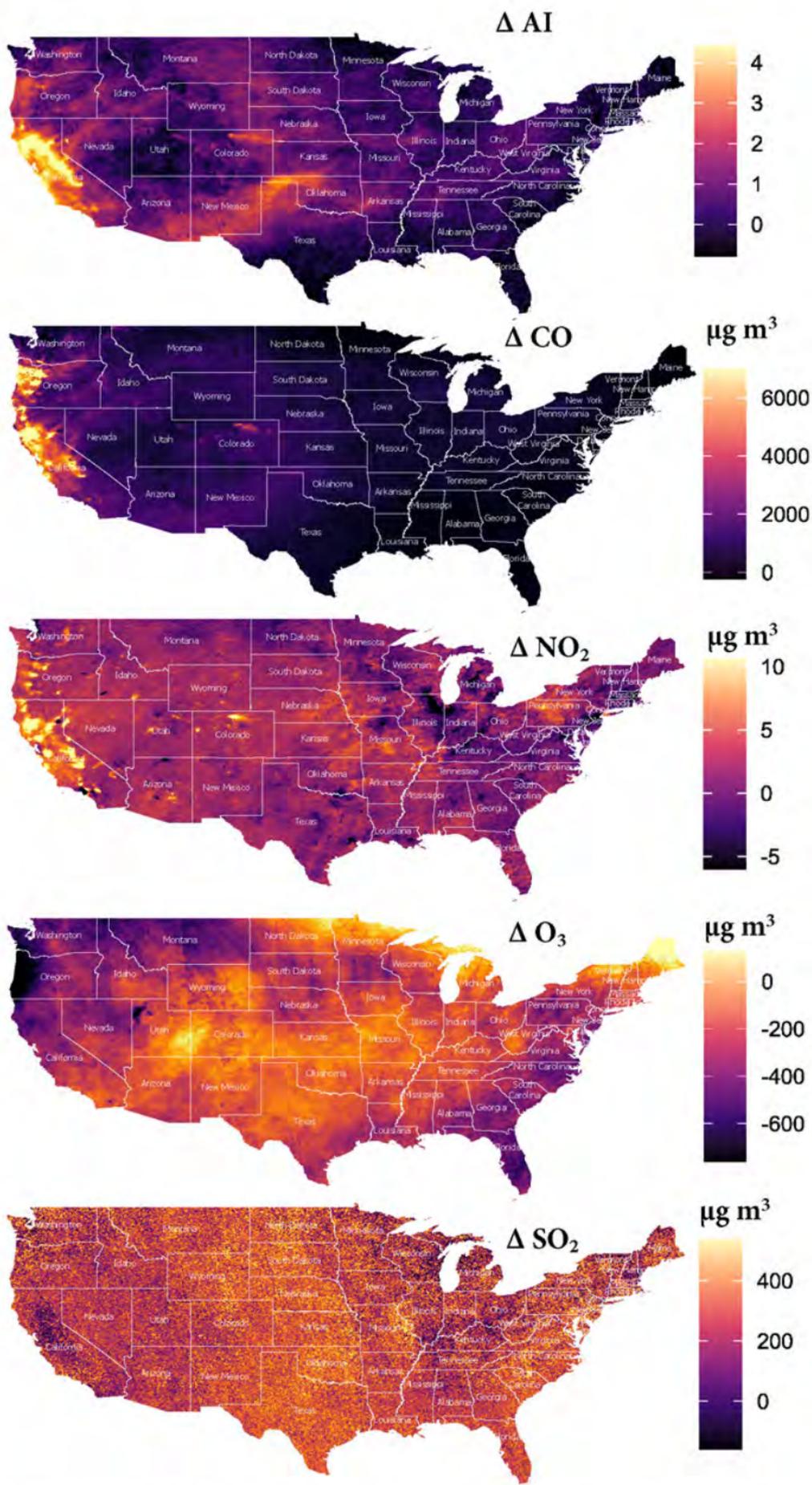

**Fig. 6** Spatial differences in air pollution concentration derived from Sentinel 5P TROPOMI data.

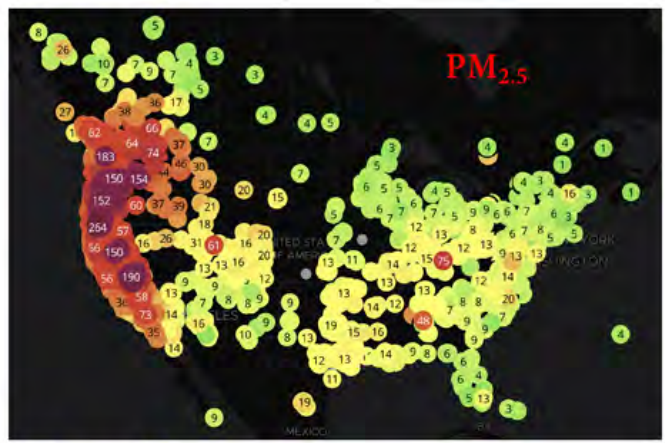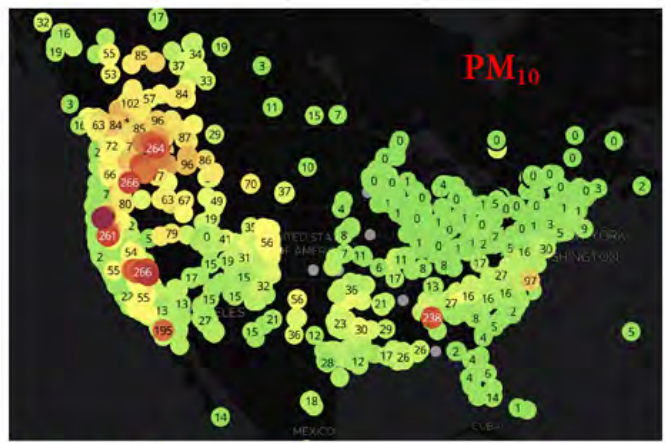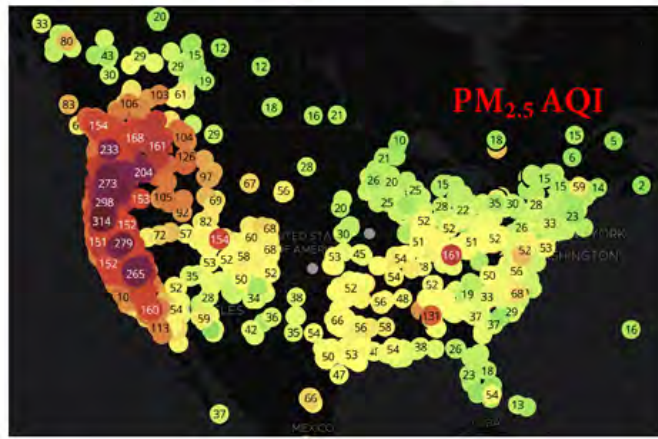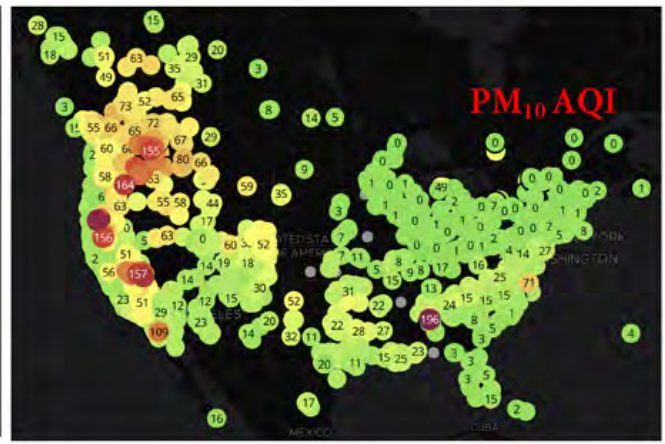

**Fig. 7** PM$_{2.5}$ and PM$_{10}$ concentration and AQI during fire occurance period.

**Table. 1** Forest fire incidents in California state during 2013 to 2020.

| Year | Area Burned (Acres) | Incidents | Loss of Life | Structures Damaged |
|---|---|---|---|---|
| 2020 | **3,472,947** | 7,882 | 25 | 6546 |
| 2019 | 259,823 | 7860 | 3 | 732 |
| 2018 | **1,975,086** | 7948 | 100 | 24,226 |
| 2017 | 1,548,429 | 9270 | 47 | 10,280 |
| 2016 | 669,534 | 6954 | 6 | 1274 |
| 2015 | 880,899 | 8283 | 7 | 3159 |
| 2014 | 625,540 | 7233 | 2 | 471 |
| 2013 | 601,625 | 9907 | 1 | 456 |

**Table. 2** Change statistics in air pollutants, i.e. CO (ppm), $PM_{25}$ and $PM_{10}$ (µg/m$^3$) during the study period (15 August to 15 September in 2019 and 2020), derived from in-situ monitored data.

| Pollutants | Year | Washington | California | Oregon |
|---|---|---|---|---|
| **CO** | **2019** | 2.34 | 0.29 | 3.16 |
|  | **2020** | 815.11 | 13.46 | 1150.26 |
|  | **Δ CO** | 812.76 | 13.17 | 1147.10 |
| **PM$_{25}$** | **2019** | 4.29 | 7.60 | 5.40 |
|  | **2020** | 50.12 | 38.87 | 47.39 |
|  | **Δ PM$_{25}$** | 45.83 | 31.27 | 41.99 |
| **PM$_{10}$** | **2019** | 18.44 | 29.40 | 23.28 |
|  | **2020** | 106.91 | 64.44 | 86.03 |
|  | **Δ PM$_{10}$** | 88.47 | 35.04 | 62.75 |

**Table. 3** Change statistics (µg/m$^3$) of air pollutants during the study period (15 August to 15 September in 2019 and 2020), derived from Sentinel 5P TROPOMI data.

| Pollutants | Year | Washington | California | Oregon |
|---|---|---|---|---|
| **AI** | **2019** | -1.26 | -0.94 | -1.18 |
| | **2020** | -0.91 | 0.44 | -0.53 |
| | **Δ AI** | -0.35 | 1.38 | -0.65 |
| **CO** | **2019** | 924.84 | 855.46 | 872.67 |
| | **2020** | 1186.80 | 1673.06 | 1571.55 |
| | **Δ CO** | 261.96 | 817.60 | 698.88 |
| **NO$_2$** | **2019** | 6.18 | 6.29 | 5.86 |
| | **2020** | 6.03 | 6.83 | 5.97 |
| | **Δ NO$_2$** | -0.16 | 0.54 | 0.11 |
| **O$_3$** | **2019** | 12357.50 | 12216.90 | 12280.26 |
| | **2020** | 11878.22 | 11831.38 | 11802.57 |
| | **Δ O$_3$** | -479.28 | -385.52 | -477.69 |
| **SO$_2$** | **2019** | 29.77 | 23.42 | 17.53 |
| | **2020** | 19.35 | 7.59 | 6.53 |
| | **Δ SO$_2$** | -10.42 | -15.83 | -11.00 |

**Table. 4** Population weighted pollution concentration of CO (ppm), PM$_{25}$ (µg/m$^3$) and PM$_{10}$ (µg/m$^3$) during the study period (15 August to 15 September 2019 and 2020).

| **Pollutants** | **Year** | **Washington** | **California** | **Oregon** |
|---|---|---|---|---|
| **CO** | 2019 | 1.72 | 0.29 | 6.73 |
| | 2020 | 566.99 | 0.96 | 2575.70 |
| | Δ CO | 565.27 | 0.68 | 2568.96 |
| | Δ CO (%) | 32823.19 | 234.95 | 38168.07 |
| **PM$_{2.5}$** | 2019 | 4.77 | 9.06 | 5.04 |
| | 2020 | 41.02 | 28.90 | 52.14 |
| | Δ PM$_{2.5}$ | 36.24 | 19.84 | 47.10 |
| | Δ PM$_{2.5}$ (%) | 759.13 | 218.99 | 935.39 |
| **PM$_{10}$** | 2019 | 17.57 | 30.88 | 21.22 |
| | 2020 | 102.77 | 68.04 | 93.32 |
| | Δ PM$_{10}$ | 85.20 | 37.16 | 72.10 |
| | Δ PM$_{10}$ (%) | 484.85 | 120.33 | 339.76 |